\documentclass[preprint,a4paper]{aastex}
\usepackage{graphics,epsf}
\usepackage{amsmath}                
\usepackage{amsfonts}               
\usepackage{amssymb}                
\usepackage{epsfig}                 
\usepackage{rotating}

\def \km{~\rm{km}}

\def \G{~\rm{G}}

\def \yr{~\rm{yr}}
\def \pc{~\rm{pc}}

\begin{document}

\title{TYPE IA SUPERNOVAE FROM VERY LONG DELAYED EXPLOSION OF CORE$-$WD MERGER}

\author{Marjan Ilkov\altaffilmark{1} and Noam Soker\altaffilmark{1}}

\altaffiltext{1}{Department of Physics, Technion -- Israel Institute of Technology, Haifa
32000, Israel;  marjan@tx.technion.ac.il; soker@physics.technion.ac.il}

\begin{abstract}
We study the spinning down time scale of rapidly rotating white dwarfs (WDs) in the frame
of the core-degenerate (CD) scenario for type Ia supernovae (SNe Ia).
In the CD scenario the Chandrasekhar or super-Chandrasekhar mass WD is formed at the termination
of the common envelope phase or during the planetary nebula phase, from a merger of a WD companion
with the hot core of a massive asymptotic giant branch star.
In the CD scenario the rapidly rotating WD is formed shortly after the stellar formation episode, and the delay from
stellar formation to explosion is basically determined by the spin-down time of the rapidly rotating merger remnant.
We find that gravitational radiation is inefficient in spinning down WDs, while the magneto-dipole radiation
torque can lead to delay times that are required to explain SNe Ia.
To explain the delay-time-distribution of SNe Ia the merger remnants distribution should be
${dN}/{d \log (B \sin \delta)} \approx {\rm constant}$, for $10^6 \G \la B \sin \delta \la 10^8 \G$
where $B$ is the dipole magnetic value and $\delta$ the angle between the magnetic dipole axis and rotation axis.
\end{abstract}

\section{INTODUCTION}
\label{sec:intro}

 Type-Ia supernovae (SNe Ia) are widely thought to be the thermonuclear detonations of carbon-oxygen white dwarfs (WDs)
whose masses are near the Chandrasekhar limit \citep{HoyleFowler1960}.
The route to achieve this mass and the conditions for explosion are in dispute
(see reviews by, e.g., \citealt{Livio2001} and \citealt{Hillebrandt2000}).
Two main routes are discussed in the literature.
In the single degenerate (SD) scenario (e.g., \citealt{Whelan1973}; \citealt{Nomoto1982}; \citealt{Han2004})
a WD grows in mass through accretion from a non-degenerate stellar companion.
In the double degenerate (DD) scenario (\citealt{Webbink1984}; \citealt {Iben1984};
see \citealt{vanKerkwijk2010} for a recent paper on sub-Chandrasekhar mass remnants)
two WDs merge after losing energy and angular momentum through the radiation of
gravitational waves.
Observations and theoretical studies cannot teach us yet whether both models for SNe Ia can work,
only one of them, or none (see recent summary by \citealt{Maoz2010}).

The observations and theoretical results that favor and disfavor each scenario for the progenitors of SNe Ia
are summarized by, e.g., \citet{Livio2001} and more recently by \citet{Howell2011}.
The main problem for the DD scenario (see review by \citealt{Howell2011})
is that in many cases an off-center carbon ignition occurs (e.g., \citealt{SaioNomoto2004})
leading to accretion induced collapse (AIC) where a neutron star (NS) is formed rather than a SNe Ia.
\cite{Yoon2007} raised the possibility that in a merger process where the more massive WD is hot,
off-axis ignition of carbon is less likely to occur.
The reason is that a hot WD is larger, such that its potential well is shallower
and the peak temperature of the destructed WD (the lighter WD) accreted material is lower.
Hence, in such a case the supercritical-mass remnant is more likely to ignite carbon in the center
at a later time, leading to a SN Ia.
Namely, the merger remnant becomes a rapidly rotating massive WD, that can collapse
only after it loses sufficient angular momentum.

Motivated by the possibility raised by \cite{Yoon2007}, \citet{KashiSoker2011} suggested recently that some SNe Ia
might originate from a double degenerate merger that occurs at the end of the common envelope (CE) phase,
rather than a long time after the CE phase as in the canonical DD scenario.
Practically, the merger is of a companion WD with the core of an AGB star.
{{{ The merger of a WD with the core of an AGB star was proposed by \citet{Sparks1974}.
They consider the outcome to be a SN explosion that forms a NS, e.g., type II SN explosion.
More relevant to this paper is the suggestion of \citet{Livio2003} that the merger of the WD with the AGB core
leads to a SN Ia that occurs at the end of the CE phase or shortly after. In that case hydrogen lines
will be detected. As \citet{Livio2003} aim was to explain the presence of hydrogen in a SN Ia,
they did not consider a delayed explosion. In this paper we discuss delayed explosion.
\citet{Livio2003} noted also that for such a a merger to occur the AGB star should be massive
(also \citealt{KashiSoker2011}).
Tout  et al. (2008) consider a merger of a WD with a core of an AGB star to explain the formation of massive rotating
WDs with strong magnetic fields. \citet{Wickramasinghe2000} commented that such WDs might be more likely to form SNe Ia. }}}

\begin{figure}
   \includegraphics[scale=0.7]{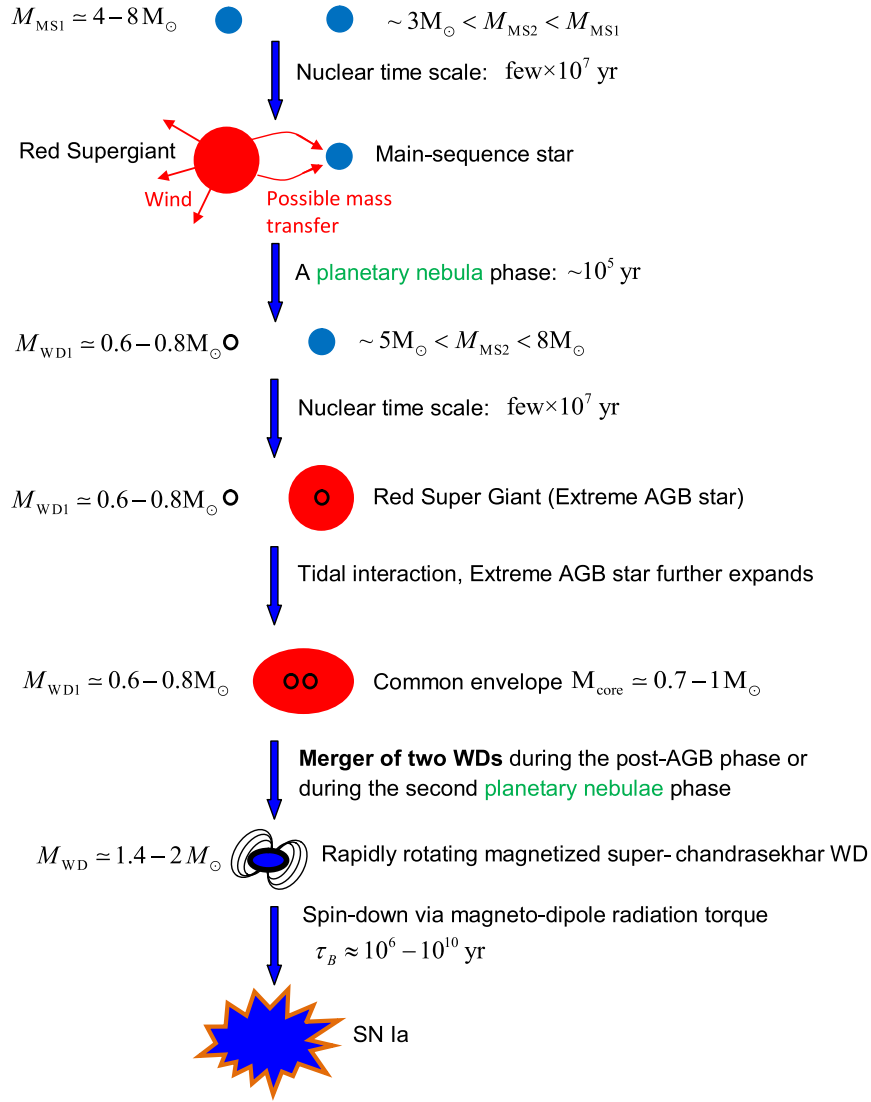}
        \caption{A schematic summary of the core-degenerate (CD) scenario for SNe Ia \citep{KashiSoker2011}.
        This paper deals with the spin-down process.  }
   \label{fig:fig1}
     \end{figure}

Either the companion WD is destructed and forms a disk around the hot core, or the more massive core is destructed
(because it is hot and large) and forms an accretion disk around the WD. In the later case the lower mass WD has
a shallower gravitational potential, hence the temperature in the accretion disk does not reach ignition temperature.
\citet{KashiSoker2011} termed this the core-degenerate (CD) scenario.
In the CD scenario the merger occurs in a CE with a massive AGB star, or shortly ($\la 10^5 \yr$) after
the CE phase, i.e. during the planetary nebula phase.
Because of its rapid rotation the super-Chandrasekhar WD does not explode (e.g., \citealt{Anand1965};
\citealt{Ostriker1968}; \citealt{Uenishi2003}; \citealt{Yoon2005}).
It will explode long after the merger, only after it has spun down to allow explosion.
The CD scenario, including findings from this paper, is summarized schematically in Figure \ref{fig:fig1}.
This scenario is completely different from the model of an explosion inside a CE with a low mass evolved star as
proposed by \citet{Hachisuetal1989} that was subsequently criticized by \citet{Applegate1991}.

There are three key ingredients in the CD scenario, in addition to the common condition that the remnant mass be
$\ga 1.4 M_\odot$.
(1) The hot core is more massive than the companion cold WD. In a future paper we will conduct a population synthesis to
find the expected number of such systems.
(2) The merger should occur while the core is still large, hence hot. This limits the merger to occur within $\sim 10^5 \yr$
after the common envelope phase. \citet{KashiSoker2011} showed that this condition can be met when the AGB star is massive,
and some of the ejected CE gas falls back.
(3) The delay between merger and explosion should be up to $\sim 10^{10} \yr$ if this scenario is to account
for SNe Ia in old-stellar populations, in addition to SNe Ia in young stellar populations.
This is the subject of this paper.

In section \ref{sec:gradiation} we examine the spinning-down time by gravitational waves.
Contrary to previous claims, we find that this time is very long, and might not be relevant to spin down WDs, despite being very
efficient in spinning down neutron stars (NS).
In section \ref{sec:long} we show that it is quite plausible that magneto-dipole radiation torque can lead to a long
delay of the explosion, up to $\sim10^{10} \yr$.
In section \ref{sec:distribution} we constrain the properties of the core-WD merger product if the CD scenario is
to account for a large fraction of SNe Ia.
Our discussion and summary are in section \ref{sec:summary}.

\section{THE LIMITED ROLE OF GRAVITATIONAL RADIATION}
\label{sec:gradiation}

As done by many authors (e.g., \citealt{Andersson99}, \citealt{Yoon2004}, \citealt{Yoon2005}, \citealt{Yoon2007};
\citealt{Justham2011}; \citealt{DiStefano2011}) we consider the emission of gravitational waves due to r-modes in
rotating WD progenitors of SNe Ia, but contrary to most of these and
other works, we will find this mechanism to be inefficient.
{{{{ We note that \citet{Piersanti2003a, Piersanti2003b} considered gravitational waves from
WDs that have rotational to binding energy ratio of $T/W=0.14$, and therefore are highly deformed.
The WDs we consider cannot reach values of $T/W=0.14$ \citep{Yoon2005}. }}}}

We follow the derivation given by \citet{Lindblom1998}, where the assumptions and approximation can
be found.
We will take the mode $l=m=2$, which is the most significant one.
The velocity perturbation of the r mode is
\begin{equation}
\delta  \overrightarrow{v} = \alpha R \Omega \left(\frac{r}{R} \right)^l \overrightarrow{Y}_{ll}^B e^{i \omega t}
\label{eq:deltav}
\end{equation}
where $\overrightarrow{Y}_{ll}^B$ is the magnetic type vector spherical harmonic, the frequency is given
by
\begin{equation}
\omega = - \frac{(l-1)(l+2)}{l+1} \Omega,
\label{eq:ome1}
\end{equation}
$\Omega$ and $R$ are the stellar angular velocity and radius, respectively, and $\alpha$ is the amplitude of the mode.
{From} equations 14-17 in \citet{Lindblom1998} we derive the power, $W_{\rm GW}$, of the gravitational radiation (for $l=m=2$,
and written in a form that can be traced to their formulae)
\begin{equation}
W_{\rm GW} = \frac{32 \pi G}{(5!!)^2c^7} \left(\frac{4}{3} \right)^6
\left( \int_0^R \rho(r) r^6 dr \right)^2  R^{-2} \Omega^8 \alpha^2.
\label{eq:power1}
\end{equation}
We define the structural constant $\beta$ by
\begin{equation}
\int_0^R \rho(r) r^6 dr = \frac {3} {28 \pi} \beta M R^4 ,
\label{eq:beta}
\end{equation}
where $\beta=1$ for a constant density sphere of mass $M$ and radius $R$.
We also define the ratio of the angular velocity to the break-up angular velocity (the Keplerian angular velocity
on the stellar equator $\Omega_{\rm Kep}$)
\begin{equation}
\tilde{\Omega} \equiv \frac{\Omega}{\Omega_{\rm Kep}} =\frac{\Omega}{(G M/R^3)^{1/2}}.
\label{eq:omega}
\end{equation}
With these definitions the gravitational radiation power becomes
\begin{equation}
W_{\rm GW} = \frac {2^{13}}{7^2 \times 5^2 \times 3^6 \pi} \frac{G^5}{c^7} M^6 R^{-6} \tilde{\Omega}^8   \beta^2 \alpha^2.
\label{eq:power2}
\end{equation}

The moment of inertia is parameterized with the structural constant $\beta_I$,
\begin{equation}
I = 0.4 \beta_I M R^2,
\label{eq:betaI}
\end{equation}
where for a constant density sphere $\beta_I=1$.
We integrate for the structure of a massive WD model taken from \citet{Yoon2005}, and find
$\beta \simeq 0.12$, and $\beta_I \simeq 0.27$.
Assuming that all the power of the gravitational waves comes from the rotational kinetic energy,
the spin-down timescale of the WD is given by integrating the equation $d(I \Omega^2/2)/dt=W_{\rm GW}$, where
a solid body rotation is assumed here.
This gives the timescale for spinning down from an initial fast rotation ${\tilde{\Omega}_0}$ down to a critical (scaled)
angular velocity $\tilde{\Omega}_c$,
\begin{equation}
\begin{split}
\tau_{\rm GW} &
 =
\frac{I \Omega_c^2}{6 W_{\rm GW}(\Omega_c)} \left[1-\left(\frac{\tilde{\Omega}_0}{\tilde{\Omega}_c}\right)^{-6}\right]
=10^{11}
\left( \frac{\beta_I / \beta^2}{10} \right)
\left(\frac {M}{1.5 M_\odot }\right)^{-4}
\\  
& \times
\left(\frac {R}{4000 \km }\right)^{5}
\left(\frac {\tilde{\Omega}_c}{0.7} \right)^{-6}
\left(\frac {\alpha}{10^{-2}}\right)^{-2}
\left[1-\left(\frac{\tilde{\Omega}_0}{\tilde{\Omega}_c}\right)^{-6}\right]
\yr.
\end{split}
\label{eq:taus}
\end{equation}

In spinning-down rapidly rotating NSs, that have a much smaller radius than WDs, gravitational radiation
emitted by the r-modes is very efficient.
For the gravitational radiation to play any role in spinning-down WDs the amplitude of the r-modes must become
$\alpha \ga 0.1$. This is quite large.
Here we raise the possibility that magnetic fields limit the growth of the r-mode to a degree
that makes the gravitational radiation insignificant in spinning down WDs.
\citet{Cuofano2010} find that as the magnetic fields in NSs become stronger, the
lower limit on the rotation frequency for the development of r-modes instability becomes larger.
In other cases the magnetic field limits the amplitude of the r-modes.
In the calculations of \citet{Cuofano2010} an equipartition between the magnetic pressure
and the mode energy density in the outer region is achieved where the density is about one per cent of
the average density $\rho \sim 0.01 \rho_{\rm av}$.
Using the same approximate equipartition, $(1/2) \rho (\alpha \Omega_{\rm rot} R)^2 \approx B^2/8 \pi$,
we obtain
\begin{equation}
\begin{split}
B \approx   10^{10}
\left(\frac{\alpha}{10^{-2}}\right)
\left(\frac {\tilde{\Omega}_c}{0.7} \right)
\left(\frac{R}{4000 \km }\right)^{-2}
\left(\frac{M}{1.5 M_\odot }\right) \G .
\end{split}
\label{eq:bfield1}
\end{equation}
This is a huge magnetic field for a WD. Before reaching this value the magnetic field will dissipate, and is
likely to prevent the growth of the r-mode amplitude to a value of $\alpha \sim 0.01$.
Our conclusion is that most likely gravitational waves from r-modes are not significant in spinning down WDs.

\section{VERY LONG DELAYED EXPLOSION BY MAGNETIC BREAKING}
\label{sec:long}

A misalignment between the magnetic axis and the rotation axis leads to a spin-down process due to
magneto-dipole radiation torque that is commonly used for pulsars (e.g., \citealt{Contopoulos2006}, \citealt{Benacquista2003}).
This mechanism was used for WDs by \citet{Benacquista2003}, with the following expression for the spin-down time
from an initial fast rotation ${\tilde{\Omega}_0}$ down to a critical angular velocity $\tilde{\Omega}_c$,
\begin{equation}
\begin{split}
\tau_{\rm B} & \simeq
\frac{I c^3}{ B^2 R^6 \Omega_{\rm c}^{2}}
\left[1-\left(\frac{\tilde{\Omega}_0}{\tilde{\Omega}_c}\right)^{-2}\right] (\sin \delta)^{-2}
\approx  10^{8}
\left(\frac{B}{10^8 \G}\right)^{-2}
\left(\frac{\tilde\Omega_{\rm c}}{0.7 \Omega_{\rm Kep}}\right)^{-2}
\\&
\times
\left(\frac{R}{4000 \km }\right)^{-1}
\left(\frac{\sin \delta}{0.1}\right)^{-2}
\left(\frac{\beta_I}{0.3}\right)
\left[1-\left(\frac{\tilde{\Omega}_0}{\tilde{\Omega}_c}\right)^{-2}\right]
\yr,
\end{split}
\label{eq:taub}
\end{equation}
where $\delta$ is the angle between the magnetic axis and the rotation axis,
and $\beta_I$ is defined in equation (\ref{eq:betaI}).
The scaling of ${\tilde{\Omega}_c}$ at which the WD explodes is based on the results of
\citet{Yoon2005} for WD in the lower range ($1.4-1.5 M_\odot$; also \citealt{Ostriker1968}).
The angular velocity distribution is not considered here in detail; only in calculating the total angular momentum
a solid body rotation is assumed. In a future paper this assumption will be relaxed. However, we do note that
magnetic fields might enforce a solid body rotation in the WD \citep{Piro2008}.

{{{  Stabilizing rapidly rotating super-Chandrasekhar WDs is a delicate matter (e.g., \citealt{Yoon2004},
\citealt{Chen2009}, and \citealt{Hachisuetal2011} in the single degenerate scenario).
The strong magnetic fields required in the present model most likely will enforce a rigid rotation within a short
time scale due the WD being a perfect conductor. The critical mass of rigidly rotating WDs is $1.48 M_\odot$
(\citealt{Yoon2004} and references therein). This implies that in the case of a strong dipole filed
that exists in most of the their interior, WDs more massive than $1.48 M_\odot$
will explode in a relatively short time. The exact time will be estimated in a future study.
The similarity of most SN Ia suggests that their progenitors indeed come from a narrow mass range.
This is $\sim 1.4-1.48 M_\odot$ in the CD scenario.
}}}

Equations (\ref{eq:taus}), (\ref{eq:bfield1}), and (\ref{eq:taub}) bring us to consider the following scenario.
Massive WDs, $M_{\rm WD} \ga 1.5 M_\odot$ that can explode (or collapse) with higher angular momentum,
do so on a typical time scale of $<10^9 \yr$; most within a much sorter time.
WDs in the lower mass domain ($M_{\rm WD} \simeq 1.4-1.48 M_\odot$) that explode with lower angular momentum
and that on average have weaker magnetic fields might survive for a time of $>10^9 \yr$ from their formation
in a merger process.
The exact time of explosion depends strongly on the magnetic field of the WD and the inclination angle.
We cannot constrain the value of these parameters and the evolution of the magnetic field during
the spin-down process, as, for example, r-modes might amplify the magnetic field.
However, in the next section we crudely constrain the distribution of the relevant parameters.

\section{DELAY TIME DISTRIBUTION}
\label{sec:distribution}

In the previous two sections we have studied the evolution of a massive rapidly rotating WD that is a merger product
of two WDs, or an AGB core with a WD, termed CD scenario for SNe Ia.
The rapidly rotating WD does not collapse because of the centrifugal forces.
Once it spins down below the critical angular velocity $\Omega_c$, it becomes unstable and explodes as a SN Ia
(\citealt{Yoon2005}).
We now check the possibility that most of the SN Ia are formed by the CD scenario, and
use observations to constrain the properties of these SN Ia progenitors .

Equation (\ref{eq:taub}) for the spin-down time depends on 4 physical variables.
The radius $R$ and the critical angular velocity $ {\tilde{\Omega}_c}$ will not differ much from one
WD to another, as we are mainly interested in WDs of masses $\sim 1.5 M_\odot$ that are rapidly rotating.
The magnetic field and the inclination angle can vary by more than two orders of magnitude between different WDs.
For that, we write the spin-down time from equation (\ref{eq:taub}) in the form
\begin{equation}
\tau_{\rm B} = F({\tilde{\Omega}_c}, R) \eta^{-2}
\label{eq:taub2}
\end{equation}
where we defined what we term the magnetic-dipole parameter
\begin {equation}
\eta \equiv B \sin \delta.
\label{eq:eta}
\end{equation}

We now try to use observations to estimate the distribution of WDs that are the remnants of core-degenerate mergers
with respect to $\eta$, under the assumption that most (all) SNe Ia are formed through this channel.
Maoz et al. (2011; see also \citealt{Graur2011}, \citealt{Maozetal2010}, \citealt{Ruiter2010}) summarize the
SN Ia rate versus time $-$ the delay time distribution (DTD) $-$ as found from observations
\begin {equation}
\frac{dN}{dt} \propto t^{\epsilon},
\label{eq:dndt1}
\end{equation}
with $\epsilon \simeq -1$, and where $t$ is the time since the star formation event that formed the
binary system that later became the SN progenitor.
The last equation can be written as
\begin {equation}
\frac{dN}{dt} = \frac{dN}{d\eta}\frac{d\eta}{dt}=t^\epsilon.
\label{eq:dndt2}
\end{equation}
{From} equation (\ref{eq:taub2}) with $\tau_{\rm B} \rightarrow t$, we have
\begin {equation}
\frac{d\eta}{dt} \propto t^{-\frac{3}{2}},
\label{eq:detadt1}
\end{equation}
that when substituted into equation (\ref{eq:dndt2}) gives
\begin{equation}
\frac{dN}{d\eta} \propto  t^{\frac{3}{2} + \epsilon} \propto \eta^{-3 - 2 \epsilon} \sim \eta^{-1}.
\label{eq:dNdeta1}
\end{equation}

We recall that this result holds only for the massive WDs that were formed from the merger of two lower mass WDs
during or shortly after a common envelope phase under the assumption that most SNe Ia are formed via the CD channel.
Equation (\ref{eq:dNdeta1}) can be also written explicitly as
\begin{equation}
\frac{dN}{d \log (B \sin \delta)} =  {\rm constant} \qquad {\rm for} \qquad  10^6 \G \la B \sin \delta \la 10^8 \G,
\label{eq:dNdeta2}
\end{equation}
where the range of $\eta = B \sin \delta$ comes from the range of $ 10^7 \la t \la 10^{10} \yr$
of the observed SNe rate \citep{Maozetal2011} and equation (\ref{eq:taub}).
{{{ We note that magnetic fields in massive WDs might be $ \ga 10\%$ of their initial value
at age of $10^{10} \yr$ \citep{Muslimov1995}.  If true, then the decay of the magnetic field should be taken into account
for long delay times, but it does not pose a problem to the model.

Let us compare this results with observations.
\citet{Wickramasinghe2000} review the subject of magnetic WDs. Magnetic WDs have
magnetic fields in the range $3 \times 10^4 - 10^9 \G$. This overlaps with the value assumed here.
Although many of them are called rapid rotators because they are
much faster rotators than most WDs, they are much slower than the spin rate required here.
It is quite possible that super-Chandrasekhar magnetic WDs can be stabilized even if the
surface spinning rate is lower than that assumed here,
if a differential rotation is maintained inside the WD (\citealt{Yoon2005}).

Another encouraging observational result is that magnetic WDs tend to be more massive that non-magnetic WDs
\citep{Wickramasinghe2000}.
Indeed, out of the 16 magnetic WDs with well determined mass reported by \citet{Wickramasinghe2000}, two have
masses of $M>1.3 M_\odot$. This fraction is further discussed in the next section. }}}

\section{DISCUSSION AND SUMMARY}
\label{sec:summary}

In this paper we examined the delay time from merger to explosion of the core-degenerate (CD) scenario for SNe Ia.
A key ingredient in this scenario is that the merger occurs while the more massive degenerate component
is the core of an AGB star or a planetary nebulae, e.g., at the end of the common envelope
phase or shortly after
{{{ Earlier studies of the merger of a WD with the core of an AGB star are \citet{Sparks1974} as a way to make a neutron star,
\citet{Livio2003} as an evolutionary route for SNe Ia with hydrogen lines, \citet{Tout2008} as a way to make magnetic WDs,
and \citet{KashiSoker2011} in relation to SNe Ia with long delay times as studied here. }}}

Such a merger where the more massive WD is still hot, hence larger, is more likely to avoid an early off-center ignition
of carbon \citep{Yoon2007}; an early off-center ignition leads to a NeMgO WD instead of to a SN Ia.
The problem of off-center carbon ignition during merger of cold WDs is the most severe problem of the double
degenerate (DD) scenario for SNe Ia  \citep{Howell2011}. The CD scenario avoids this problem.
The remnant of the core-degenerate merger is a rapidly rotating WD. If its mass is super-critical it will explode
after spinning-down to a critical angular velocity $\Omega_c$ \citep{Yoon2005},
{{{ and not necessarily in a very short time delay as in the scenario of \citet{Livio2003} where one is expect to
observe hydrogen lines. }}}

In section \ref{sec:gradiation} we examined the spinning-down time by the commonly used gravitational waves mechanism.
Although this mechanism is know to be very efficient in spinning down neutron stars, we found that this
mechanism is not efficient in spinning down WDs (eq. \ref{eq:taus}).
It can be efficient only if the amplitude of the r-modes becomes
$\alpha \ga 0.1$. This is a very large amplitude and the modes are likely to be damped by magnetic fields before reaching
this value (eq. \ref{eq:bfield1}).
On the other hand, in section \ref{sec:long} we estimated the spinning-down time due to magneto-dipole radiation
torque (eq. \ref{eq:taub}), and found it to operate over the required time scale of $\sim 10^6-10^{10} \yr$
for plausible parameters of the massive rapidly rotating WD remnant of the core-degenerate merger.

If the CD scenario for SNe Ia is to explain most (all) SNe Ia, it should account for the delay time distribution (DTD),
i.e., the SNe Ia rate versus time since their progenitor formation.
Using the observations as summarized by \cite{Maoz2010}, we derive a crude distribution of the dipole parameter
$\eta \equiv B \sin \delta$, where $B$ is the magnetic field and $\delta$ the inclination angle between the magnetic dipole axis
and the rotation axis. This is given in equation (\ref{eq:dNdeta2}).

Before the CD scenario can be accepted as one of the channels to form SNe Ia more detailed calculations should be done.
We end with a summary of these calculations, and with some possible promising properties of this scenario.
\begin{enumerate}
\item The claim of \cite{Yoon2007} that when the massive WD in a merger is hot (in the CD scenario it is the core of an AGB star
or of a planetary nebula), the off-center ignition of carbon is likely to be avoided, should be studied in more detail.
This study should include the process where the more massive core is destructed (because of its larger radius), and the
lower mass WD becomes the core of the remnant.
\item A population synthesis study is required to determine the number of core-WD mergers that lead to Chandrasekhar
and super-Chandrasekhar mass remnants.
This is a subject of a future paper. The first step was carried by \cite{KashiSoker2011} who argued
that the core-WD merger can be a common outcome of the common envelope phase with massive AGB stars, $\sim 5-8 M_\odot$.
{{{ Tout et al. (2008) estimated that about three times as many common envelope events lead to a merged
core as to a cataclysmic variable. This is a high fraction.
\citet{Wickramasinghe2000} estimated that $\sim 5\%$ of all WDs are magnetic WDs.
Even if a small fraction of these merger products
lead to SNe Ia, the CD scenario can account for a large faction of SNe Ia. The exact numbers will be derived in
a forthcoming paper. }}}
The large fraction of massive stars that are in binary systems gives us hope that indeed the number of such systems might be adequate
to explain most (all) SNe Ia. We note here that SNe Ia from WDs masses much above the Chandrasekhar mass, as observed
in some cases, e.g. \citet{Howell2001}, are accounted for in the CD scenario.
\item A detailed study of the spin-down time of the merger remnant due to the magneto-dipole radiation torque is required.
Here we showed that this can be a promising mechanism.
The evolution of the magnetic field with time should be examined as well.
\item It is expected that less massive remnant (but still super-Chandrasekhar) will take longer time on average to spin
down and explode. This might account for the finding that SNe Ia
in older populations are less luminous (e.g., \citealt{Howell2001}). This should be studied after the
magneto-dipole radiation torque process is better explored.
\item Rapidly rotating WDs have shallow density profile near the equatorial plane \citep{Yoon2005}.
Shallow density profile favors incomplete silicon burning that is required by observations \citep{Pakmor2010}.
This process deserved to be studied.
\item
Eventually, rapidly rotating massive WDs should be found.
{{{ Claims for massive rapidly rotating and strongly magnetized WDs have already been made,
e.g. see discussion by \citet{Malheiro2011} and \citet{Wickramasinghe2000}. }}}
We can crudely estimate the expected number of
such WDs if the CD scenario account for most of SNe Ia, in the following way (Maoz, D., private communication 2011).
The SN Ia rate per unit stellar mass in Sbc galaxies is $\sim 1.5 \times 10^{-13} \yr^{-1} M_\odot^{-1}$ \citep{Li2011}.
{From} the local stellar density of $0.085 M_\odot \pc^{-3}$ \citep{McMillan2011} and the local WD density of
$0.0046 \pc^{-3}$ \citep{Harris2006}, a number of stellar mass per WD of $18.5 M_\odot {\rm WD}^{-1}$ in the
solar neighborhood is obtained.
If these WDs are to explode within $\sim 10^{10} \yr$ with a time time delay distribution of $t^{-1}$ in the range
$10^7-10^{10} \yr$, then their
average age is $\sim 1.5 \times 10^9 \yr$, and the fraction of SN Ia progenitors should be $\sim 0.4\%$ of all WDs.
Namely, in a sample of $10^{4}$ WDs in the solar neighborhood the number of rapidly rotating massive WDs should be $\sim 40$.
{{{ As $\sim 5\%$ of all WDs are magnetic \citep{Wickramasinghe2000}, it is expected that $\sim 10\%$ of the magnetic WDs
will be massive enough to become SNe Ia. This is a prediction of this model.
Indeed, out of the 16 magnetic WDs with well determined mass reported by \citet{Wickramasinghe2000}, two have
masses close to the Chandrasekhar limit, $M>1.3 M_\odot$. }}}
The study of large samples of WDs is the subject of a future paper.

\end{enumerate}

 {{{ We thank an anonymous referee for helpful comments. }}}
 This research was supported by the Asher Fund for Space Research
at the Technion, and the Israel Science foundation.


\label{lastpage}

\end{document}